\documentstyle[aps,preprint]{revtex}
\begin{document}

\title{Collectivity Embedded in Complex Spectra\\
of Finite Interacting Fermi Systems: Nuclear Example}

\author{S. Dro\.zd\.z$^{1,2}$, S. Nishizaki$^{3}$, J. Speth$^{1}$
and M. W\'ojcik$^{1,2}$}
\address{
$^{1}$Institut f\"ur Kernphysik, Forschungszentrum J\"ulich,\\
D-52425 J\"ulich, Germany \\
$^{2}$ Institute of Nuclear Physics, PL-31-342 Krak\'ow, Poland\\
$^{3}$ Faculty of Humanities and Social Sciences, Iwate University, \\
Ueda 3-18-34, Morioka 020, Japan }
\date{\today}
\maketitle

\begin{abstract}
The mechanism of collectivity coexisting with chaos in a finite system of
strongly interacting fermions is investigated. The complex spectra are
represented in the basis of two-particle two-hole states describing the
nuclear double-charge exchange modes in $^{48}$Ca.
An example of $J^{\pi}=0^-$ excitations shows that the residual interaction,
which generically implies chaotic behavior, under certain specific and well
identified conditions may create strong transitions, even much stronger than
those corresponding to a pure mean-field picture.
Such an effect results from correlations among the off-diagonal
matrix elements, is connected with locally reduced density of states
and a local minimum in the information entropy.

\end{abstract}



\newpage

\section{Introduction}

The concept of the random matrix theory (RMT)~\cite{Brod} proves very fruitful
in approaching complex quantum systems and in addressing the question
of how classical chaos manifests itself on the quantum level.
Chaos is essentially a generic property of complex systems such as atomic
nuclei~\cite{Zel1}, many electron atoms~\cite{Fla1}, molecules~\cite{Zimm}
or disordered mesoscopic systems~\cite{Alts} and this finds evidence
in a broad applicability of RMT to describe level fluctuations~\cite{Guhr}.
Even many aspects of quantum chromodynamics are consistent with
chiral RMT~\cite{Verb}.
Similarly, however, as in most physically interesting cases where
classical chaos is not just a hard billiard-type chaos, the pure
RMT cannot account for the full richness of quantum phenomena
connected with complexity. As an example one can mention the sign
correlations~\cite{Fran} for parity nonconserving effects~\cite{Fla2}
in compound nuclei, even though it was the physics of compound nuclei
which lead Wigner~\cite{Wign} to postulate the Gaussian orthogonal
ensemble (GOE) of random matrices as an appropriate global frame.
Explicit microscopic approaches in terms of the full shell
model diagonalization, either in nuclear~\cite{Zel1,Kusn} or atomic
physics~\cite{Fla1}, show perfect agreement with GOE when looking at
the local level fluctations measured in terms of the nearest neighbor
spacing distribution and the $\Delta_3$ statistics,
but significant deviations take place on
the level of wave functions. This originates from the two-body nature
of interaction which reduces the number of independent
parameters and preserves certain correlations among the matrix elements.
In order to account for this type of correlations a two-body random
interaction model has been introduced~\cite{Fren} and its statistical
properties investigated in detail~\cite{Fla3}.
Still however, such models may not properly account for correlations
which originate from geometry of a problem and which, in some cases,
may turn out significant.

Another characteristics connected with complexity, even more interesting
and important from the practical point of view, is collectivity.
It means a cooperation, and thus the coupling between the different degrees
of freedom in order to generate a coherent signal in response to an external
perturbation. Consequently, even though the real collectivity implies a highly
ordered behavior it involves effects beyond the mean-field -- the most regular
part~\cite{Zel2} of the many-body Hamiltonian.
At the same time the effects beyond the mean field are responsible for the
GOE fluctuation properties. Therefore, in a sense, these two seemingly
contradictory phenomena, chaos and collectivity, may have to go in parallel.
Also on classical level collectivity is a nonlinear cooperative effect
which results from the coupling between different degrees of freedom.

In general, the shell model type approaches are based on diagonalization
of the full many-body Hamiltonian in the basis spanned by all possible
n-particle -- n-hole (npnh) configurations generated by the mean-field.
For practical reasons, especially when large energy intervals are involved,
as for instance in the case of nuclear giant resonances,
one truncates this hierarchy of configurations up to 2p2h~\cite{Dro1}.
Interestingly, due to a sufficiently large density of states
relative to the strength of the residual interaction~\cite{Jacq},
local level fluctuations characteristic of GOE appear~\cite{Dro2}
to take place for the nuclear Hamiltonian acting already in the space
of 2p2h states and this is a crucial
element for an appropriate description of the giant resonace decay
properties~\cite{Dro3}. The giant resonaces are however excited by
one-body operators which directly probe
the 1p1h components of the nuclear wave function.
The 2p2h states only form the background which determines a decay-law.
There exist, however, very interesting physical processes,
represented by two-body (two-phonon) external operators, which directly couple
the ground state to the space of 2p2h states.
In view of the above mentioned local GOE fluctuations giving evidence
for a significant amount of chaotic dynamics already in the 2p2h space,
the question of a possible coherent response or collectivity under
such conditions is a very intriguing one and of interest
for many branches of physics.

\section{Model}

We start from the Hamiltonian which, in second quantized form,
reads as
\begin{equation}
\hat H=\sum_i\epsilon_i a_i^{\dag} a_i+{1\over 4}\sum_{ij,kl}v_{ij,kl}
a_i^{\dag} a_j^{\dag} a_la_k.
\label{eq:H}
\end{equation}
The first term denotes the mean field
while the second term is the residual interaction
with antisymmetrized matrix elements $v_{ij,kl}$.
Diagonalizing this Hamiltonian in the subspace of 2p2h states
\begin{equation}
\vert 2 \rangle \equiv
a_{p_1}^{\dag} a_{p_2}^{\dag}  a_{h_2}a_{h_1}\vert 0 \rangle
\label{eq:2}
\end{equation}
yields the eigenenergies $E_n$ and the corresponding eigenvectors
$\vert n \rangle = \Sigma_2 c^n_2 \vert 2 \rangle$. For realistic
nuclear interactions the spectral fluctuations of $\{E_n\}$
typically coincide with those of GOE~\cite{Dro1}.

The general form of matrix elements for the two-body residual
interaction ${\cal V}$ between 2p2h-states is given by
\begin{eqnarray}
& &{\cal V}_{p_1p_2h_1h_2,p_1'p_2'h_1'h_2'}
\nonumber\\
&=&\delta_{p_1p_1'}\delta_{p_2p_2'}v_{h_1'h_2'h_1h_2}
 +\delta_{h_1h_1'}\delta_{h_2h_2'}v_{p_1p_2p_1'p_2'}\\
&+&a(p_1,p_2)a(h_1,h_2)a(p_1',p_2')a(h_1',h_2')
 \delta_{p_2p_2'}\delta_{h_2h_2'}v_{p_1h_1'h_1p_1'},
\nonumber
\label{eq:V22}
\end{eqnarray}
where $a(i,j)$ denotes the antisymmetrizer between $i$ and $j$.
The consecutive terms in this expression are responsible for hole-hole,
particle-particle and particle-hole interactions, respectively, while
the remaining pair of states in each case are spectators represented
by the $\delta_{ij}$ functions. These functions set a significant
fraction of the matrix elements to zero which may lead to correlations.
Fig.~1 illustrates the corresponding structure in diagrammatical
representation. Further correlations may originate from
the fact that many nonzero matrix elements relate to each other only
by the geometrical factors due to the angular momentum coupling algebra.

In response to an external field ${\hat F}_{\alpha}$ a state
\begin{equation}
\vert F_{\alpha} \rangle \equiv  {\hat F}_{\alpha} \vert 0 \rangle =
\sum_n \langle n \vert {\hat F}_{\alpha} \vert 0 \rangle \vert n \rangle
\label{eq:F}
\end{equation}
is excited. The two-phonon operator ${\hat F}_{\alpha}$ can be
represented as
\begin{equation}
{\hat F}_{\alpha}=[{\hat f}_{\beta}\otimes{\hat f}_{\gamma}]_{\alpha},
\label{eq:ph}
\end{equation}
where ${\hat f}_{\beta}$ and ${\hat f}_{\gamma}$ denote the single-phonon
operators whose quantum numbers $\beta$ and $\gamma$ are coupled
to form $\alpha$.
The state $\vert F_{\alpha} \rangle$ determines the strength function
\begin{equation}
S_{F_{\alpha}}(E) = \sum_n S_{F_{\alpha}}(n) \delta (E-E_n),
\label{eq:S}
\end{equation}
where 
\begin{equation}
S_{F_{\alpha}}(n) =
\vert\langle n \vert{\hat F}_{\alpha}\vert 0 \rangle\vert^2.
\label{eq:sn0}
\end{equation}
In the unperturbed basis of states $\vert 2 \rangle$
the transition strength $S_{F_{\alpha}}(n)$ to the state
$\vert n \rangle$ can be expressed as
\begin{eqnarray}
& &S_{F_{\alpha}}(n)
\nonumber\\
&=&\sum_2 \vert c^n_2 \vert^2
\vert \langle 2 \vert {\hat F}_{\alpha} \vert 0 \rangle \vert^2
+ \sum_{2\neq2'} c^{n*}_2 c^n_{2'}
\langle 0 \vert {\hat F}_{\alpha}^{\dag} \vert 2'\rangle
\langle 2 \vert {\hat F}_{\alpha} \vert 0 \rangle
\nonumber\\
&=& S^d_{F_{\alpha}}(n) + S^{od}_{F_{\alpha}}(n).
\label{eq:Sn}
\end{eqnarray}
The second equality defines the diagonal
$(S^d_{F_{\alpha}}(n))$ and off-diagonal
$(S^{od}_{F_{\alpha}}(n))$ contributions to the transition strength
at energy $E_n$. The second component includes many more terms
and it is this component which potentially is able to induce
collectivity,
i.e. a strong transition to energy $E_n$. Two elements are however
required: (i) a state $\vert n \rangle$ must involve sufficiently many
expansion coefficients $c^n_2$ over the unperturbed states
$\vert 2 \rangle$ which carry the strength
($\langle 2 \vert {\hat F}_{\alpha} \vert 0 \rangle \neq 0$) and
this is equivalent to at least local mixing,
but at the same time (ii) sign correlations among these expansion
coefficients should take place so that the different terms do not cancel
out.

Optimal circumstances for the second condition to be fulfilled read:
\begin{equation}
c^n_2 \sim \langle 0 \vert {\hat F}_{\alpha} \vert 2 \rangle.
\label{eq:sim}
\end{equation}
This may occur if the interaction matrix elements can be represented by a sum
of separable terms ${\bf Q}^{\nu}$ of the multipole-multipole type
\begin{equation}
{\cal V}_{ij,kl}=\sum_{\nu=1}^M Q^{\nu}_{ij} Q^{\nu}_{kl}
\label{eq:sep}
\end{equation}
with $Q^{\nu}_{ij} \sim \langle i \vert {\hat f}_{\nu} \vert j \rangle$.
The success of the Brown-Bolsterli schematic model~\cite{BB}
in indicating the mechanism of collectivity on the 1p1h level
points to an approximate validity of such a representation
and its formal justification comes from the multipole expansion of the
residual interaction. The structure of the Hamiltonian matrix
in the 1p1h-subspace is then usually dominated by few multipoles.
Collectivity can then be viewed as an edge effect connected with
appearance of a dominating component in the Hamiltonian matrix and the rank
$M$ of this component is significantly lower (unity in case of the
Brown-Bolsterli model) than the size of the matrix. This rank specifies
a number of the prevailing states whose expansion coefficients predominantly
are functions of ${\bf Q}^{\nu}$. In general, on the 2p2h level
a multipole structure of the interaction enters the corresponding matrix
elements in a more complicated way. However,
due to two-body nature of the nuclear interaction which reduces its 2p2h
matrix elements to combinations of the ones representing the
particle-particle, hole-hole and particle-hole interactions~\cite{Dro1}
the separability may become effective also on the 2p2h level although
conditions are expected to be more restrictive. On the other hand
the 2p2h space offers many more unperturbed transitions to form a collective
state and the net effect may still appear significant.

For quantitative discussion presented below we choose the $^{48}$Ca
nucleus, specify the mean field part of the Hamiltonian~(\ref{eq:H})
in terms of a local Woods-Saxon potential including the Coulomb
interaction and adopt the density-dependent zero-range interaction
of ref.~\cite{SW} as a residual interaction (after correcting for a misprint
in the density functional: $R_{0}=1.16 A^{1/3}$).
Since we want to inspect the higher energy region
at least three mean field shells on both sides
of the Fermi surface have to be used to generate the unperturbed 2p2h
states as a basis for diagonalization of the full
Hamiltonian~(\ref{eq:H}).
Typically, the number of such states is very large and this kind of
calculation can be kept under full numerical control only for selected
excitations of the lowest multipolarity.
Among various nuclear excitation modes which can be considered in this
context the double-charge exchange (DCX) processes are of special
interest. These modes, excited in $({\pi}^+,{\pi}^-)$ reactions~\cite{Leit},
involve at least two nucleons within the nucleus and give rise to
a sharp peak at around 50 MeV in the forward cross section.
They are thus located in the energy region of the high density of 2p2h
states which points to the importance of coherence effects among those
states. Consequently, the present investigation may also appear helpful
in studying the mechanism of DCX reactions and in separating
the suggested~\cite{Bilg} dibaryon contribution from the conventional
effects~\cite{Kaga}.
For all these reasons we perform a systematic study of the DCX $J^{\pi}=0^-$
states. Our model space then develops N=2286 2p2h states.
There are still several possibilities of exciting such a double-phonon
mode represented by the operator ${\hat F}_{\alpha}$
out of the two single-phonons ${\hat f}_{\beta}$ and ${\hat f}_{\gamma}$
of opposite parity. For definitness we choose
\begin{equation}
{\hat f}_{\beta}= r Y_1 {\tau}_-
\label{eq:f1}
\end{equation}
and
\begin{equation}
{\hat f}_{\gamma}= r^2 [Y_2 \otimes \sigma]_{1^+} {\tau}_-.
\label{eq:f2}
\end{equation}
The first of these operators corresponds to the $1\hbar\omega$ dipole
and the second to $2\hbar\omega$ spin-quadrupole excitation.
The resulting two-phonon mode thus operates on a level of
$3\hbar\omega$ excitations.
Formulas needed to express the angular momentum coupled form
of the above one- and two-body operators can be found, for instance, in the
appendix of ref.~\cite{Buba}.

\section{Results and Discussion}

The results of calculations are presented in Fig.~2. As one can see,
including the residual interaction (part (b)) induces
a strong transition at 51.1 MeV. This transition
is stronger by almost a factor of 2 than any of the unperturbed
(part (a)) transitions even though it is shifted to a significantly
higher ($\sim$ 10 MeV) energy.
This is also a very collective transition. About $95\%$ of the
corresponding strength originates from $S^{od}_{F_{\alpha}}(n)$,
as comparison between parts (b) and (c) of Fig.~2 indicates.
This whole effect is due to particle-hole type matrix elements ((c) in Fig.~1).
Discarding diagrams (a) and (b) produces no significant difference.
The degree of mixing can be quantified, for instance,
in terms of the information entropy~\cite{Izra}
\begin{equation}
I(n)=-\Sigma_i p_i \ln p_i;~~~~~~~~p_i=\vert c^n_i \vert^2
\label{inf}
\end{equation}
of an eigenvector $\vert n \rangle$ in the basis (part (d)).
Interestingly, the system finds preferential conditions
for creating the most collective state in the energy region
of local minimum in $I(n)$.
Our following discussion is supposed to shead more light on this issue.

As shown in Fig.~3(a) our Hamiltonian matrix displays a band-like structure
with spots of the significant matrix elements inside. This together with
a nonuniform energy distribution $\rho_u(E)$ of the unperturbed 2p2h states
(Fig.~3(b)), which is a trace of the shell structure of the single
particle states, characteristic of many other mesoscopic systems~\cite{Nazm},
sizably suppresses the range of mixing and locally supports conditions for the
edge effect to occur in the energy region of the minimum in $\rho(E)$.
A comparison with Fig.~2(b) shows that the collective state is located
at about this region. Moreover, the minimum survives diagonalization
($\rho_p(E)$ in Fig.~2(c)) and all the above features are consistent with
the effective band range~\cite{Fein}
\begin{equation}
(\Delta E_i)^2 = \sum_j(H_{ii}-H_{jj})^2 H_{ij}^2 / \sum_{i \ne j} H_{ij}^2
\label{eq:band}
\end{equation}
shown in Fig.~2(d).

Further quantification of the character of mixing between the unperturbed
states is documented in Fig.~4. The distribution $P(H)$ of off-diagonal matrix
elements (a) is not Gaussian but of the following type:
\begin{equation}
P(H)=a \vert H\vert^b \exp(-\vert H\vert/c)
\label{eq:dist}
\end{equation}
This indicates the presence of the dominating multipole-multipole components
in the interaction~\cite{Zel1,Fla1}.
An interesting feature is the asymmetry between the positive and negative
valued matrix elements (see parameters in the caption to Fig.~4(a)).
The positive matrix elements are more abundant
which expresses further correlations among them
and the fact that the interaction is predominantly repulsive for the mode
considered. Significant reduction of dimensionality is also indicated
by the distribution of eigenvalues of the residual interaction.
As shown in part (b) of Fig.~4 the majority of these eigenvalues
is concentrated around zero and thus constitute approximate zero modes
of that part of the Hamiltonian. We also would like to note at this point,
without showing the results explicitly, that similar analysis on the 1p1h level
using appropriately larger model spaces (for better statistics) shows
an even larger fraction of such zero modes. This is due to the fact that
in the 1p1h-space the multipole-multipole structure of the interaction
manifests itself in a more transparent way.

Appearance of a strong transition at certain energy $E_n$
means that the structure of the Hamiltonian matrix of the residual interaction,
at least locally at around that particular energy, is governed by a component
of the type as specified by eq.~(\ref{eq:sep}) with a small number of terms
($M \ll N$) including, of course, the ones which coincide with an external
field. In a pure case a structure like this causes an energy gap between
the collective state and the remaining states. In the present case, of its
only local nature, one expects a local minimum in the density of states
in the vicinity of the collective state. Indeed, as can be seen by a careful
inspection of Fig.~2(b) versus Fig.~3(c) any stronger transition is located
in such a minimum whose range typically extends over an energy interval of the
order of 0.5 MeV. Even relatively weak transitions are asigned their own
minima. Moreover, as we have verified in certain selected cases, many other
minima in the density of states which are not occupied by the above specified
transitions turn out to be filled in by the DCX $J^{\pi}=0^-$ transitions
connected with other combinations of two one-phonon operators (for instance
${\hat f}_{\beta}= r^2 Y_2 {\tau}_-$ and
${\hat f}_{\gamma}= r [Y_1 \otimes \sigma]_{2^-} {\tau}_-$).

A reduction of the rank (real dimensionality) of the Hamiltonian matrix
evidenced above is also consistent with the observed minimum in the
information entropy (Fig.~2(d)).
Simply, in the relevant energy region there are fewer free parameters
and this sets additional constraints on the degree of mixing and thus
on the amount of chaos.
As a chaos related characteristics we take the spectral rigidity measured
in terms of the $\Delta_3$ statistics~\cite{Brod}. We find this measure more
appropriate for studying various local subtleties of mixing than the
nearest neighbor spacing (NNS) distribution because for a smaller
number of states the latter sooner becomes contaminated by strong fluctuations.
Indeed, the spectral rigidity (Fig.~5) detects differences in the level
repulsion inside the string of eigenvalues (i1) covering the first maximum in
$\rho_{p(u)}(E)$ (35.2-44.5 MeV, 400 states starting from $n=351$ up
to $n=750$) and the one (i2) covering the minimum
and thus including the collective state (44.5-52.1 MeV, 400 states from
$n=751$ to $n=1150$).
The deviation from GOE is more significant in i2 which, similarly as $I(n)$,
signals a more regular dynamics in the vicinity of the collective
state $(n=1089)$.

Conditions corresponding to the actual Hamiltonian are not the
most optimal ones from the point of view of the collectivity of our
$J^{\pi}=0^-$ DCX excitation. By multiplying the residual interaction
by a factor of $g=0.7$ we obtain a picture as shown in Fig.~6. Now the
transition located at 48.6 MeV is another factor of 2 stronger than before and,
again, all significant transitions are situated in the local
minima of $\rho(E)$ and in the overall minimum of the information entropy.
A too severe decrease of $g$ will eventually bring all the transitions to
their mean-field values. This transition is however not just linear.
Here we seem to be facing a competition of the two elements.
One is the residual interaction which must be sufficiently strong to correlate
many states but the other one is condition for the eddge effect to occur.
As a result, even for $g=0.35$ we still obtain very strong transitions,
apparently due to the fact that the interaction strength is such that
the unperturbed transitions are moved just to the absolute minimum in $\rho(E)$.

The range of values of a multiplication factor which produces
this kind of picture is rather narrow
and this feature of collectivity resembles a classical
phenomenon of the stochastic resonance~\cite{Wies}.
It is relatively easy to completely destroy so strong transitions.
By multiplying the residual interaction by a factor of $g=2.5$
(which is equivalent to increasing the density of states)
the strength distribution displays a form as shown in the lowest panel
of Fig.~6. This strength remains largely localized in energy but
the distribution of the corresponding $S_{F}(n)$ (Fig.~6) does not deviate
much from the Porter-Thomas (P-T) distribution~\cite{PT}
$P(s)=(2\pi s)^{-1/2} \exp(-s/2)$ characteristic of GOE,
even though correlations among the matrix elements are the same as before.
Interestingly, even here the larger transitions are located in their own small
minima in $\rho(E)$.
Further increase of the multiplication factor
may again produce some transitions which are more collective than those
allowed by P-T.
In particular, starting from values $\sim$4 some new strong collective
transitions appear at the upper edge of the whole spectrum.

To illustrate statistics of the transition strength
versus P-T distribution we use a measure introduced in ref.~\cite{Dro4}.
Consequently, for all the cases considered above we calculate the total number
$N$ of transitions of magnitude smaller than a given threshold value $S_{th}$,
as a function of $S_{th}$. Since the number of large components relative
to the small ones is of primary interest in the present study and in order to
set the same scale when comparing different cases we, in addition, in each case
independently, devide all the transitions $S(n)$ by the corresponding
maximum value of $S(n)$. After that $S_{max}(n)=1$ in each case.
Consistently, the RMT limit of this measure is then drawn from the
cummulative P-T distribution and this limit is indicated by the solid
line in Fig.~7. In the log-log scale this limit develops a long straight
line segment with the slope of 0.5 which reflects the dominant role of the
preexponential factor $(s^{-1/2})$ in P-T distribution at smaller transitions.
As it can be seen from this Figure, the $g=2.5$ case is very close to this
limit. But, interestingly, even $g=0.35$ tends to the same slope when probing
the region of small transitions which means that such transitions are
consistent with GOE.
Only the unperturbed case is distinct in this sense.

Finally, Fig.~8 shows the transition strength distribution of the
'constituent' single-phonon modes specified by eqs.(\ref{eq:f1})
and (\ref{eq:f2}), respectively, in their own 1p1h sectors and the same
model space of single-particle states is used. There are 28 (single) charge
exchange $J^{\pi}=1^-$ and 25 $J^{\pi}=1^+$ 1p1h states in this space.
As before, the $g$-factors reflect the strength of the residual interaction
relative to the original one. The results collected in this Figure
provide further evidence that collectivity observed on the 2p2h level
is not accidental. It can always be traced back to collectivity of the
corresponding single-phonon modes in their own subspaces. Consistently with
our previous discussion, this correspondence cannot however be expressed
simply in terms of proportionality. For instance, the single-phonon transitions
for $g=0.35$ are significantly weaker than for $g=1$ while the opposite
applies to the resulting two-phonon mode.
Recent study of ref.~\cite{Dinh}, even though based on a much simpler model,
also shows that characteristics of the two-phonon mode (double giant dipole)
are much more sensitive to detailed form of the Hamiltonian than those of
the corresponding single-phonon modes.

Taken together, a real collectivity, by which we mean a transition stronger
than those generated by the mean field, is a very subtle effect and is not
a generic property of the complex spectra. Its appearance, as it happens
for one of the components of the $J^{\pi}=0^-$ DCX excitations considered
here, involves several elements like correlations among the matrix elements,
nonuniformities in the distribution of states and a proper matching
of the interaction strength to an initial (unperturbed) location
of the transition strength relative to the scale of nonuniformities
in the distribution of states.
If present, a collective state is then located in the region of more
regular dynamics characterized by lower information entropy, more sizable
deviations from GOE of the level fluctuations and local minima in the density
of states. This later effect can thus potentially be used in experimental
studies as an extra criterion to detect collectivity.
We also would like to point out that these aspects of collectivity parallel
an analogous property hypothesised for living organisms~\cite{Kauf}
and stating that collectivity is a phenomenon occuring at the border
between chaos and regularity.


This work was supported in part by Polish KBN Grant No. 2 P03B 140 10
and by the German-Polish scientific exchange program.


\newpage
\begin{center}
{\bf FIGURE CAPTIONS}
\end{center}
{\bf Fig.~1.} Diagrammatic representation of the two-body matrix
elements in the space 2p2h states with explicit indication of the angular
momentum coupling scheme. The consequtive terms represent hole-hole,
particle-particle and particle-hole interactions, respectively.\\
{\bf Fig.~2.}~(a) The unperturbed transition-strength distribution
in $^{48}Ca$ for the $J^{\pi} = 0^-$ DCX excitation involving
the single-phonon dipole and $2\hbar\omega$ spin-quadrupole modes.
(b) The same as (a) but after including the residual interaction.
(c) $S^d_{F_{\alpha}}$ components of the transition-strength as defined
by eq.~(\ref{eq:Sn}) (notice different scale).
(d) The information entropy of the states
$\vert n \rangle$ in the unperturbed basis. The dashed line indicates
the GOE limit $(\ln(0.48N))$.\\
{\bf Fig.~3.}~(a) Structure of the Hamiltonian matrix for the
$J^{\pi}=0^-$ DCX states. The states are here labeled by energies,
ordered in ascending order and the matrix elements $H_{ik} \ge 0.1$
are indicated by the dots.
(b) density of the unperturbed 2p2h-states.
(c) density of states after the diagonalization.
(d) energy range of interaction between the unperturbed states.\\
{\bf Fig.~4.}~(a) Distribution of off-diagonal matrix elements between
the $J^{\pi}=0^-$ DCX states (histogram). The solid lines indicate fit in terms
of eq.~\ref{eq:dist} with the resulting parameters:
$a=676$, $b=-1.21$, $c=0.69$ (left) and $a=692$, $b=-1.22$, $c=0.81$ (right).
(b) Density of states corresponding to the residual interaction part
of the Hamitonian~(\ref{eq:H}).\\
{\bf Fig.~5.} Spectral rigidity $\Delta_3(L)$ for eigenvalues from the
two 400 states long intervals: $n=351-750$ (i1) and $n=751-1150$ (i2).
The long-dashed line corresponds to Poisson level distribution and the
short-dashed line to GOE.\\
{\bf Fig.~6.} Transition-strength distribution, density of states and
the information entropy for the same excitation as in Fig.~1
but the residual interaction is now multiplied by a factor of $g=0.35$
(upper part), $g=0.7$ (middle part) and $g=2.5$ (lower part), respectively.\\
{\bf Fig.~7} The total number $N$ of transitions of given strength (properly
rescaled, see text) below a threshold value $S_{th}$. The open crosses refer
to the unperturbed case ($g=0$), thick dots to $g=0.35$, open squares
to $g=0.7$, filled squares to $g=1$ and open triangles to $g=2.5$.
The solid line represents the same quantity determined from a Porter-Thomas
distribution.\\
{\bf Fig.~8} Transition-strength distribution corresponding to single-phonon
operators specified by eq.~(\ref{eq:f1}) (left) 
and by eq.~(\ref{eq:f2}) (right), calculated in the space 
of 1p1h states for various residual interaction multiplication factors $g$.

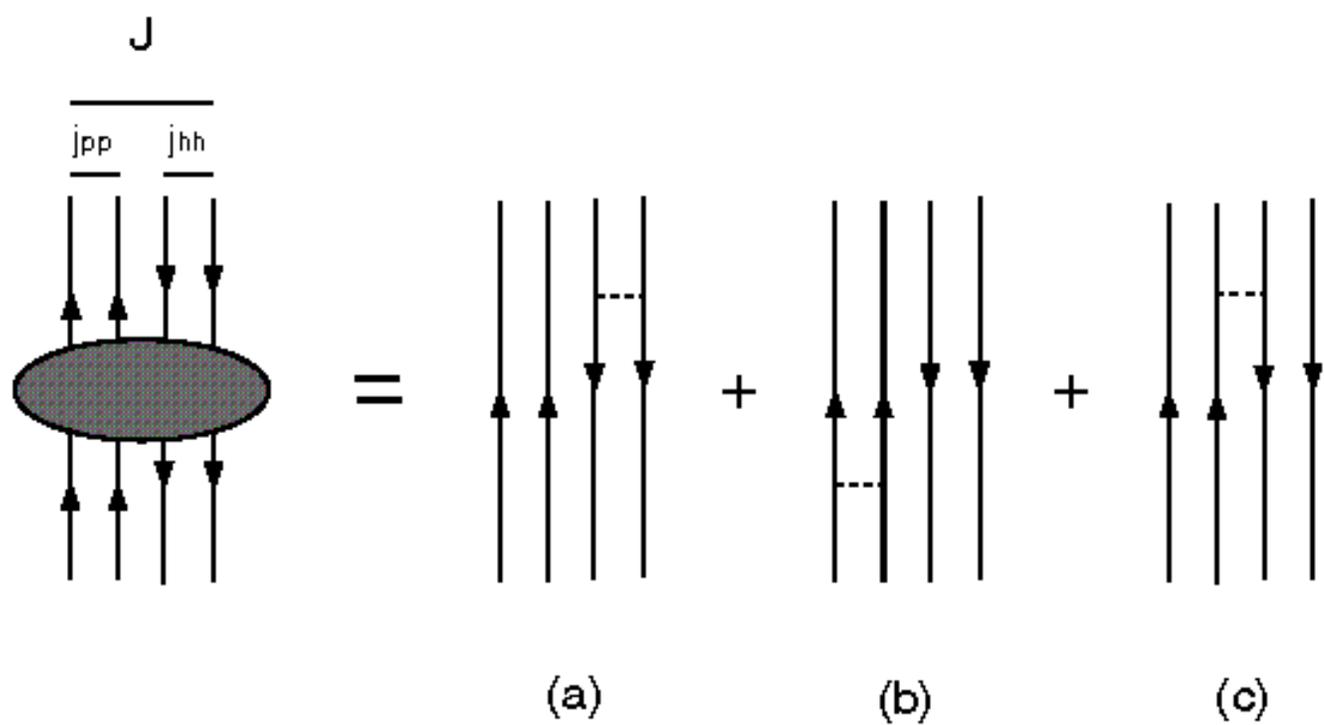
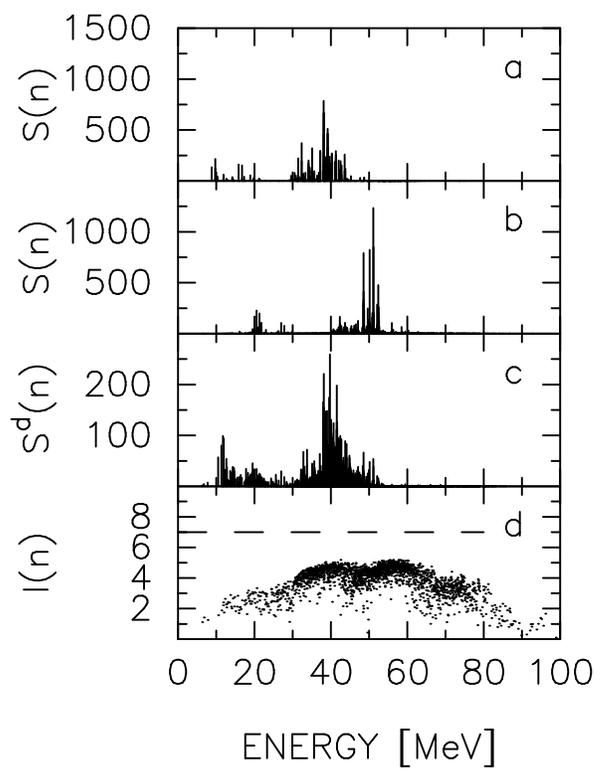
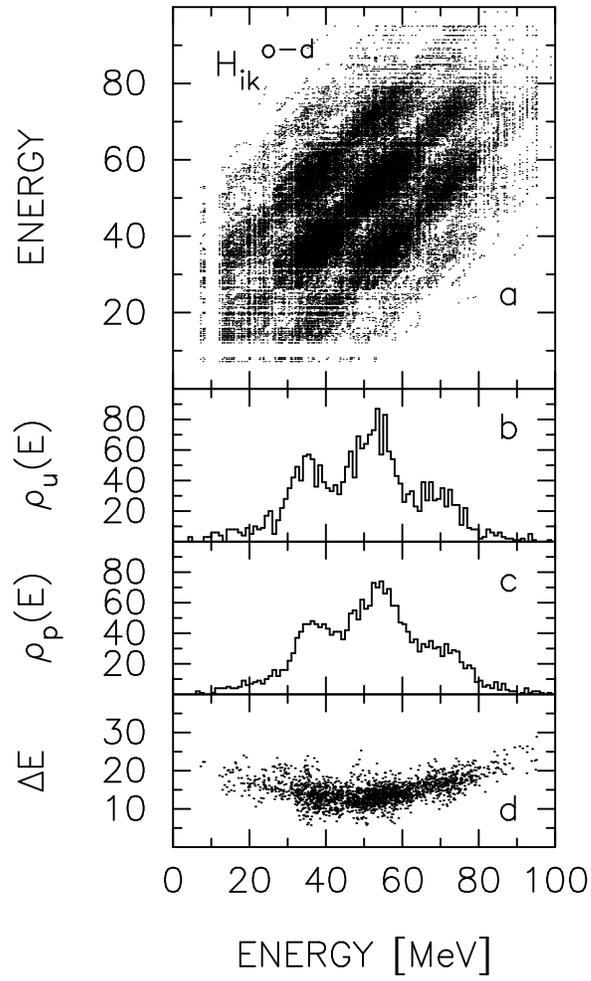
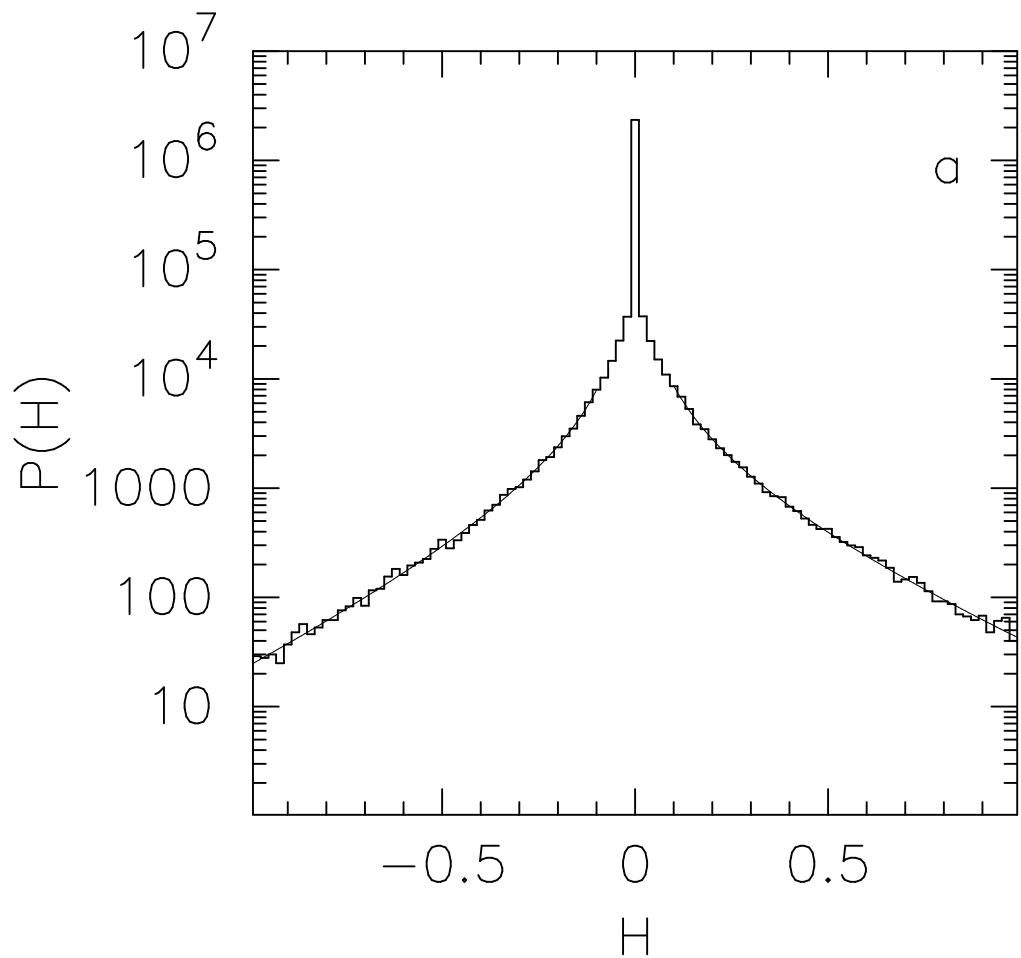
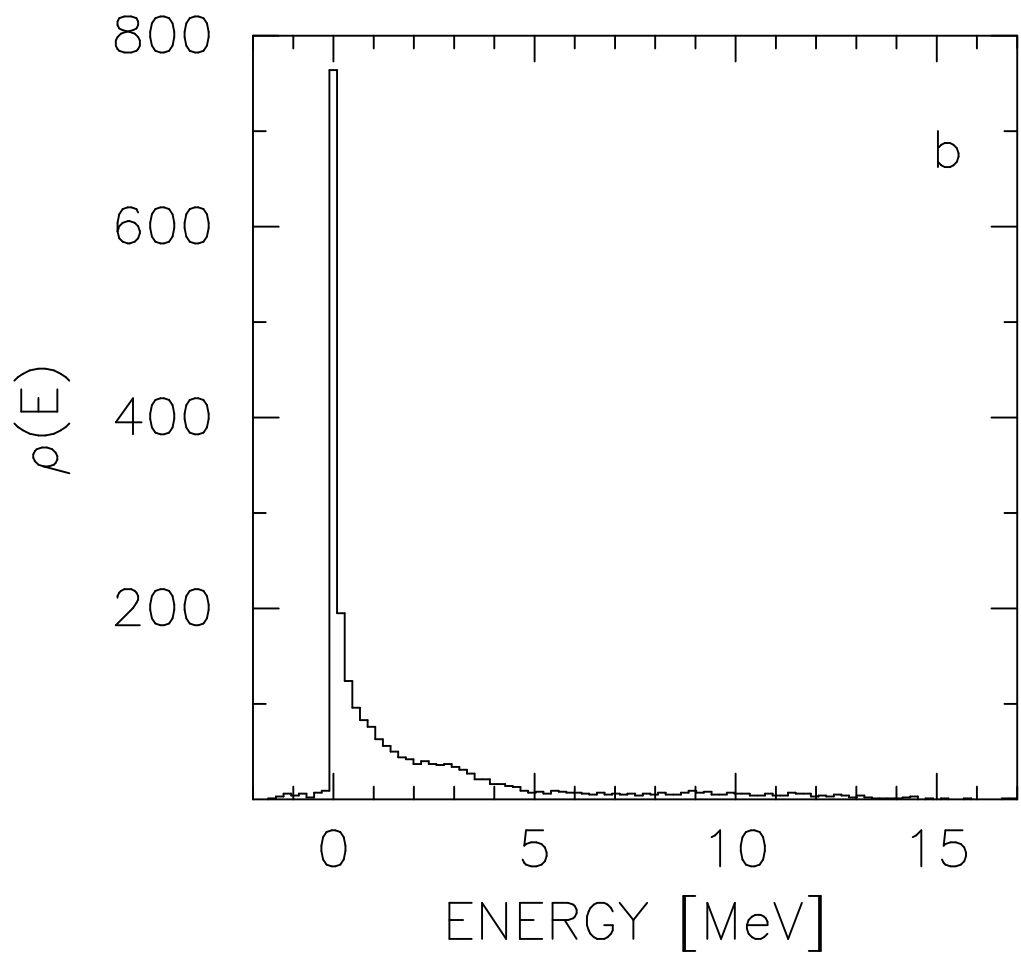
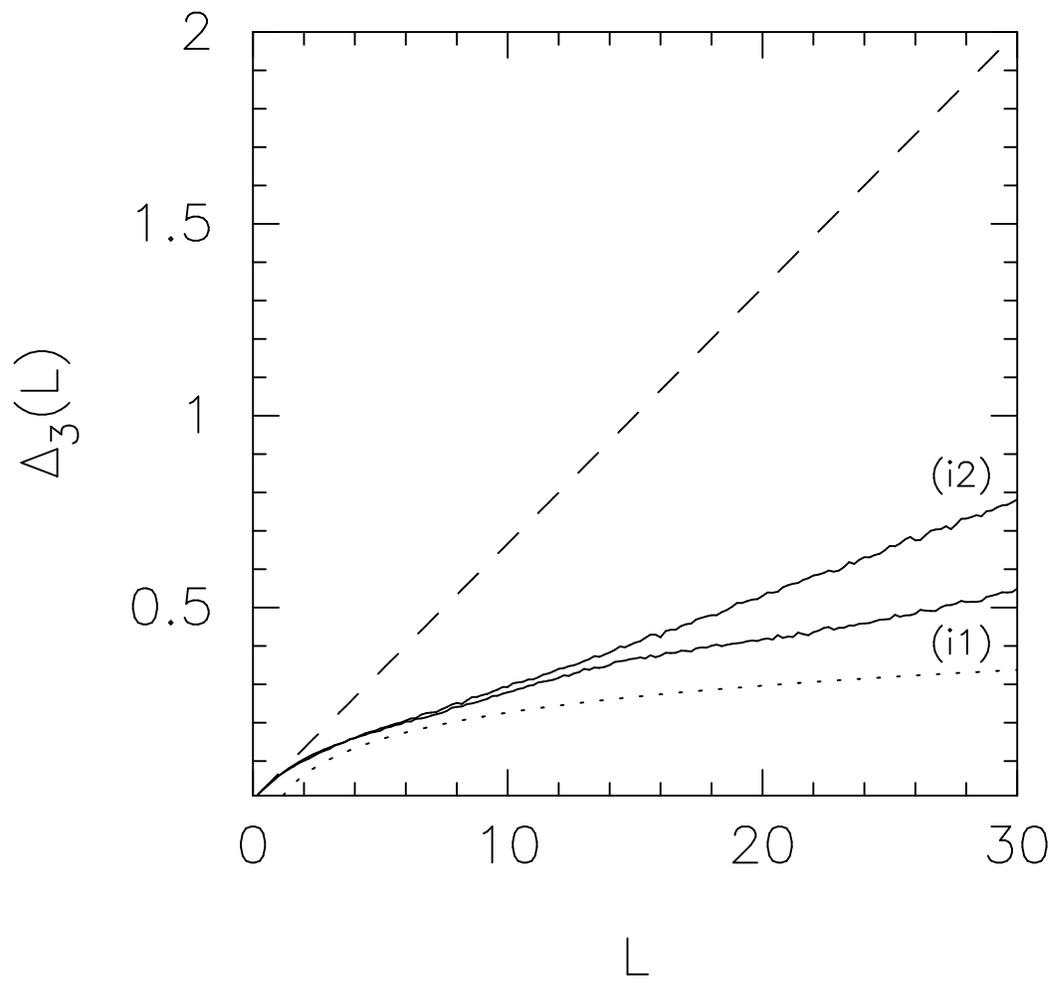
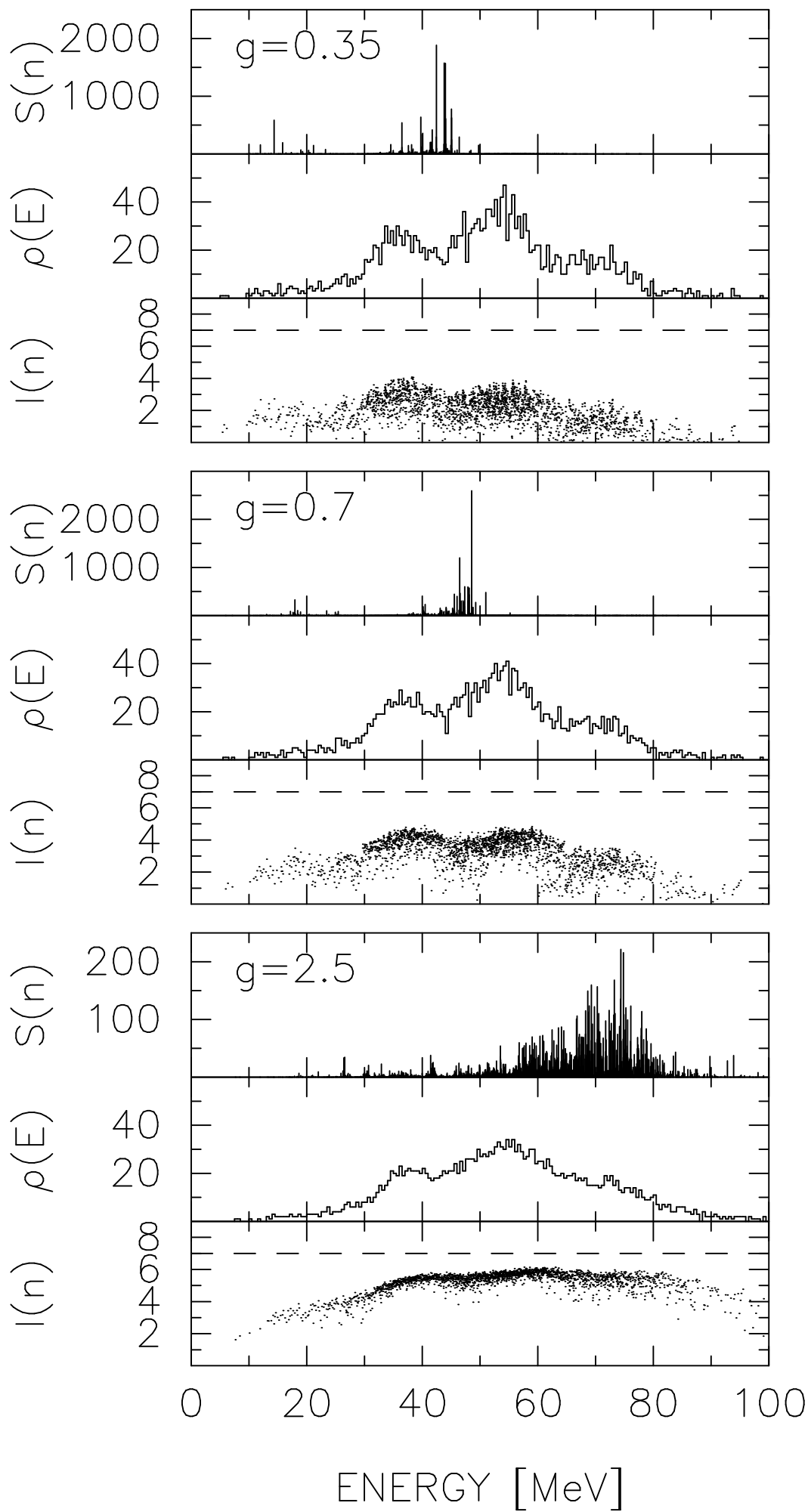
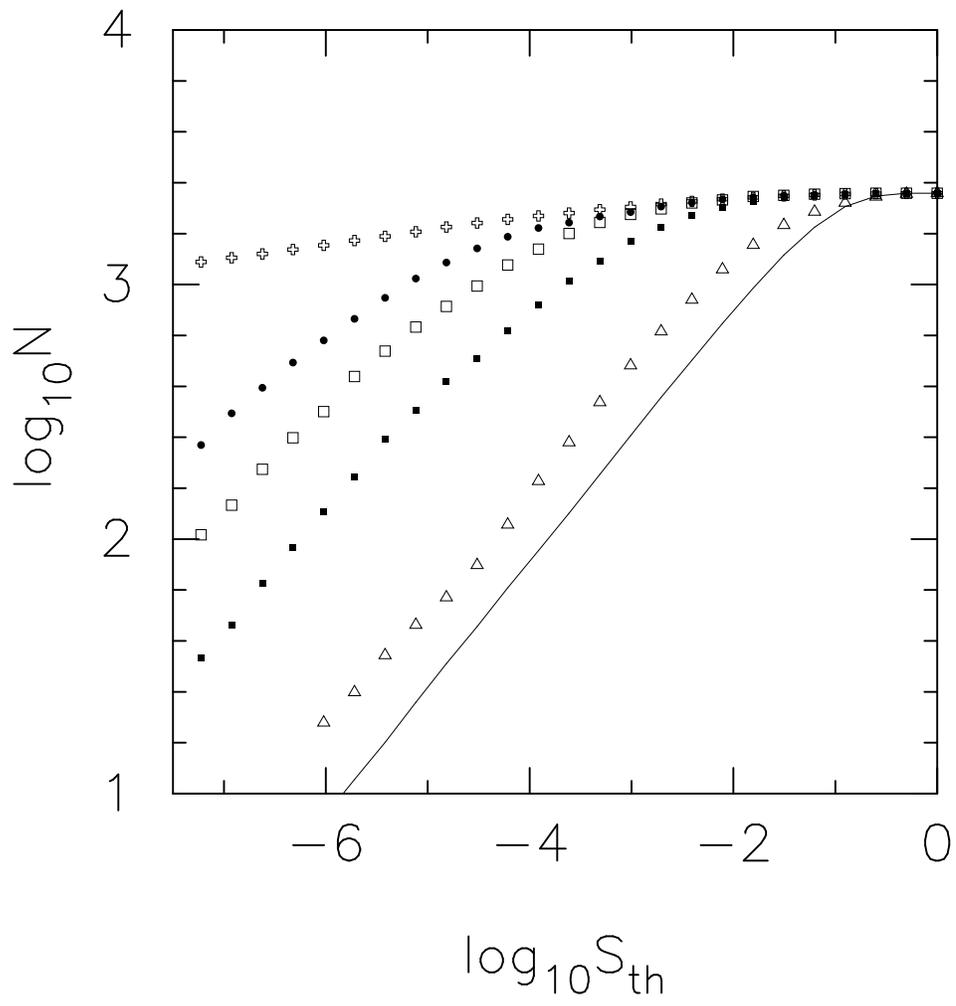
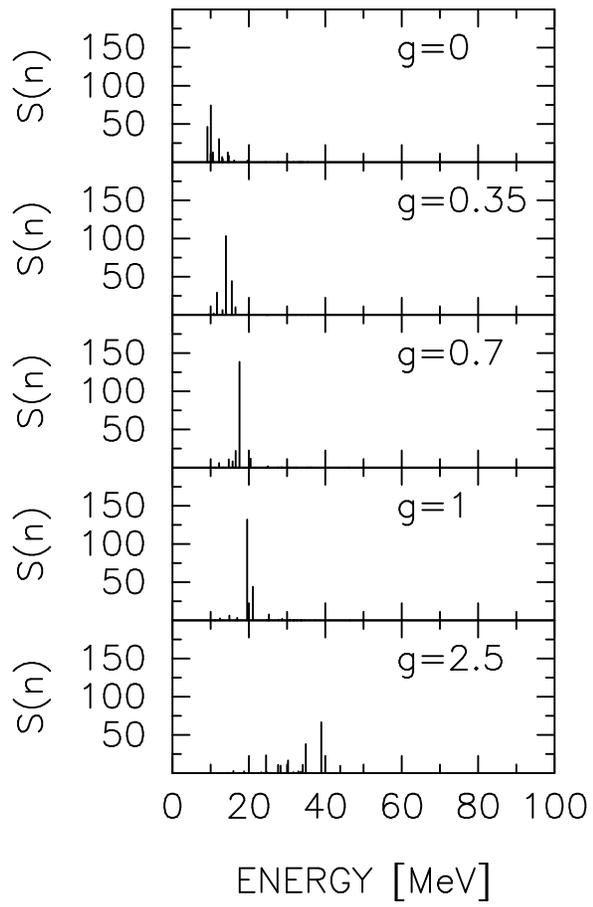
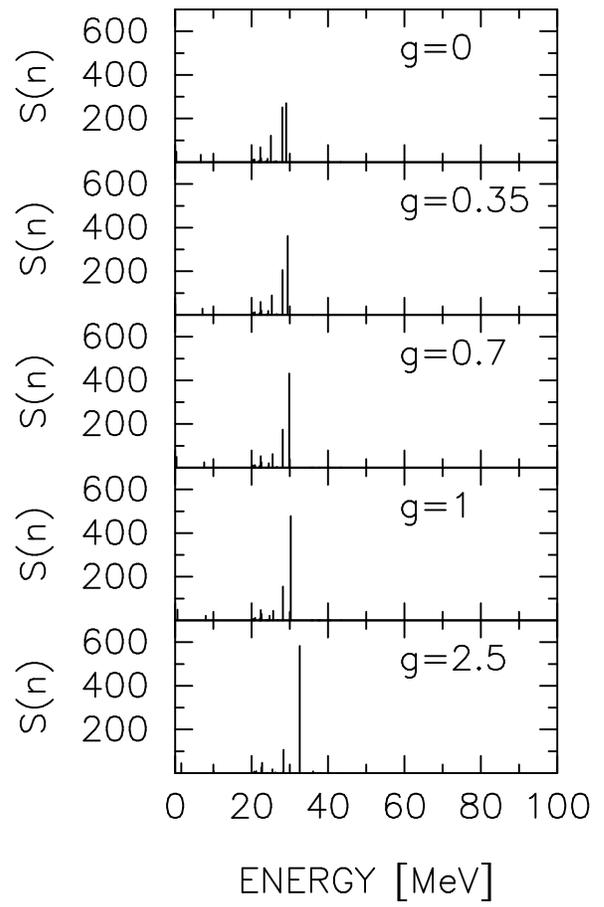
\end{document}